\documentclass[preprint2,tighten]{aastex}

\usepackage{natbib,graphicx,xspace,color} \singlespace

\shorttitle{The mass content of Abell 586} \shortauthors{Cypriano et al.}

\newcommand\eq{\begin{equation}} \newcommand\eeq{\end{equation}}
\newcommand\eqn{\begin{eqnarray}} \newcommand\eeqn{\end{eqnarray}}

\def\kms{km~s$^{-1}$\xspace}  \def\ergs{erg~s$^{-1}$\xspace}

\begin{document}


\title{Gemini and Chandra observations of Abell 586, a relaxed strong-lensing
cluster.}

\author{E. S. Cypriano\altaffilmark{1,2}, G. B. Lima Neto\altaffilmark{3},  L.
Sodr\'e Jr.\altaffilmark{3}, Jean Paul Kneib\altaffilmark{4,5} and Luis E.
Campusano\altaffilmark{6}}

\altaffiltext{1}{Southern Astrophysics Research Telescope, Casilla 603, La
Serena, Chile} \altaffiltext{2}{Laborat\'orio Nacional de Astrof\'{\i}ca, CP
21, 37500-000 Itajub\'a - MG, Brazil} \altaffiltext{3}{Universidade de S\~ao
Paulo, Instituto de Astronomia, Geof\'{\i}sica e Ci\^encias Atmosf\'ericas,
Departamento de Astronomia, Rua do Mat\~ao 1226, Cidade Universit\'aria,
05508-900, S\~ao Paulo, SP, Brazil} \altaffiltext{4}{Caltech-Astronomy,
MC105-24, Pasadena, CA 91125, USA} \altaffiltext{5}{Observatoire
Midi-Pyr\'en\'ees, CNRS-UMR5572, 14 Avenue E. Belin, 31400 Toulouse, France}
\altaffiltext{6}{Universidad de Chile, Departamento de Astronom\'{\i}a, Casilla
36-D, Santiago, Chile}

\begin{abstract} We analyze the mass content of the massive strong-lensing
cluster Abell 586 ($z = 0.17$). We use optical data (imaging and spectroscopy)
obtained with  the Gemini Multi-Object Spectrograph (GMOS) mounted on the 8-m
Gemini-North telescope, together with publicly available X-ray data taken
with the \textit{Chandra} space telescope. Employing different techniques --
velocity distribution of galaxies, weak gravitational lensing, and X-ray
spatially resolved spectroscopy -- we derive mass  and velocity dispersion
estimates from each of them. All estimates agree well with each other, within a
68\%  confidence level, indicating a velocity dispersion of 1000 -- 1250~\kms.
The projected mass distributions obtained through weak-lensing and  X-ray
emission are strikingly similar, having nearly circular geometry. We suggest
that Abell 586 is probably a  truly relaxed cluster, whose last major merger
occurred more than $\sim 4$~Gyr ago.

\end{abstract}


\section{Introduction}

In a Universe where structures are formed hierarchically, as the current
$\Lambda$CDM paradigm predicts, larger objects are formed through merging
and/or accretion of smaller systems and the most massive structures should be
the youngest. In the present era of evolution of the universe, 
massive clusters of galaxies occupy
the upper limit of the mass function of (nearly) virialized structures. This
special characteristic makes clusters privileged objects for the study of 
structure formation  and  the nature of dark matter, which is believed to play
a major role in this process \citep[e.g.][]{white78,kauffmann99}.

Cluster masses can be measured by several techniques, each one relying on
different principles and simplifying assumptions,
Consequently, their biases are also different. Two techniques -- the kinematics
of the member galaxies and the X-ray emission from the hot gas that fills the
intra-cluster medium (ICM) -- assume dynamical equilibrium. On
the other hand, gravitational lensing, in its strong and weak regime,  does
not require such an assumption, depending directly of the cluster surface mass 
\citep[see the reviews by, e.g.,][]{FM94,mellier99}.

The use of different techniques to measure mass distributions will likely give
different mass estimates. This was first demonstrated by \citet{miralda}, who
have found a systematic difference by a factor of  $\sim 2$--3 between strong
lensing and X-ray mass measurements, with the second method systematically
producing smaller values. \citet{allen} claimed to have solved this problem
after comparing results for clusters with and without cooling flows. He found
that, for the former, X-ray and lensing mass estimates tend to agree, whereas
for the latter the opposite happens. With the assumption that cooling flow
clusters are dynamically more evolved, he concluded that this discrepancy is
due to non-thermal processes (such as merger of sub-components) affecting the
ICM in the central regions of non-cooling flow clusters.

Gravitational lensing is often claimed to be a more reliable method for 
cluster mass estimation, because it does not depend on equilibrium assumptions 
about the dynamical state of the cluster. Among dynamical mass estimators, the
X-ray emission of the intra-cluster gas is preferred to galaxy dynamics, 
since the gas is a collisional component that reaches equilibrium faster than
galaxies. In other words, gas is considered a better tracer of the 
gravitational potential than galaxies \citep[see, e.g.,][Sect.~5.5]{Sarazin88}.

On the other side, there are growing evidences that lensing masses for some
clusters can be overestimated due to the presence of other massive components
along the line of sight \citep{metzeler,hoekstra}. This seems to be the case 
for the strong lensing clusters A2744 (AC118) \citep{girardi,cypriano}  and
CL0024+16 \citep{oliver,kneib03}. In these cases, distance information  is
needed (e.g., from redshift surveys)  to disentangle clearly two or more
components along the line of sight.

Therefore, the use of a multi-technique approach to investigate galaxy
clusters \citep[e.g.,][]{morphs,valtchanov,ferrari03,proust03,graham03,graham}
is ideal for the study of these objects. Not only it does allow more reliable
mass estimations, but it may also offer several hints on the physical state of
such complex and evolving systems. The goal of this paper is to apply 
this approach to investigate
the mass distribution  and the dynamical state of the cluster Abell 586.

Abell 586 is a Bautz-Morgan type I cluster, at $z = 0.17$, with Abell 
richness class 3. It presents copious X-ray emission
\citep[$L_X = 1.11 \times 10^{45} h_{50}^{-1}$ \ergs;][]{BCS},
with its peak coincident with the 
position of the brightest cluster galaxy (BCG). We have chosen this particular 
cluster for this study because of its high X-ray
luminosity and the presence of strong lensing features, both implying a high
mass content.

\citet{dahle} reported a long faint blue arc at the northwest of the cluster 
central galaxy, and our optical observations with Gemini revealed a fainter 
arc at the opposite direction with respect to the BCG.

Recent X-ray studies of this cluster \citep{Allen00, white2000}, using ROSAT
images and ASCA spectra, have found ICM temperatures ranging from 6.1 to
8.7~keV, depending on the adopted model. A de-projection analysis made by
\citet{Allen00} predicts a velocity dispersion of $1050^{+450}_{-250}$\kms for
this cluster. However, a weak-lensing mass measurement by \citet{dahle} leads 
to a much larger velocity dispersion: $\sigma = 1680^{+160}_{-170}$ \kms.

In this paper we present new optical and X-ray observations (Section 2) that
allowed us to estimate the cluster mass based on galaxy dynamics, weak and
strong lensing, and ICM temperature and surface brightness 
profiles (Section 3). In Section 4 we compare and discuss the results obtained
with the different methods and, finally, we summarize our results  and present
our main conclusions in Section 5. We adopt hereafter $H_0 = 70~ h_{70}$ 
\kms~Mpc$^{-1}$, $\Omega_M = 0.3$ and $\Omega_\Lambda=0.7$. At the distance of
Abell~586, one arcsecond corresponds to 2.9 kpc.  All quoted uncertainties
are for a confidence level of 1-$\sigma$, or 68\%, unless stated
otherwise.

\section{Observations and data reduction}

In this section we present the
optical and X-ray data used in our analysis and describe the main steps in
their reduction.

\subsection{Gemini-North optical data}

All optical observations discussed here
were obtained using the Frederick C. Gillett Telescope (Gemini North) at  Mauna
Kea operating in queue mode.  Imaging and multi-object spectroscopy were
carried out with GMOS \citep{GMOS}.  Image and spectroscopic basic reductions
(de-biasing, flat-fielding, wavelength calibration, etc.) were done in a
standard way, with the GMOS package  running under the {\sc IRAF}
environment.

\subsubsection{Imaging}

We observed the cluster Abell 586 on two occasions. The first time (period
2001B) as part of a survey for gravitational arcs in 8 clusters with X-ray 
luminosities larger than $10^{45} h_{50}^{-1}$\ergs in the BCS
catalog \citep{BCS}.

This imaging consisted of three single exposures with the g$^\prime$,
r$^\prime$ and i$^\prime$ Sloan
filters \citep{sloan}, the integration times being 300, 250 and 250 seconds,
respectively. Atmospheric conditions were nearly photometric and the seeing,
as measured by the FWHM of point sources, was $\sim$0.7\arcsec. With these
observations we detected gravitational arcs in Abell 586, that had already been
reported by \citet{dahle}.

During Gemini period 2002B, follow-up observations of the cluster were
performed, comprising deeper r$^\prime$ imaging and also multi-object
spectroscopy. The total exposure time for imaging was 20 minutes (4 $\times$
300s). These new images were taken in photometric conditions, and the seeing of
the combined image is again $\sim$0.7\arcsec.

In both cases we have used GMOS with $2\times 2$ binned pixels, leading to a
pixel size of 0.145\arcsec. The observed field of view has 5.5\arcmin\ on a
side. 

We use the program {\sc SExtractor} \citep{SEx} to build galaxy catalogs,
adopting, as detection criterion, that objects should have at least 10 
contiguous pixels with values above the background plus 1.5 times its 
dispersion. For the first run, the galaxy catalogs start to become 
incomplete (i.e., the logarithmic number counts start to departure from a 
linear behavior) at 22.5, 22.5 and 22.0 mag, for g$^\prime$, r$^\prime$ and 
i$^\prime$ bands, respectively. The magnitude completeness limit
for the second run data is r$^\prime = 24.5$ magnitudes.

\subsubsection{Spectroscopy} 

The GMOS spectroscopic observations were done using a 400 line mm$^{-1}$ red
optimized grating, with a central wavelength of 7000\AA. This configuration
resulted in spectra with a resolution of 7\AA, or nearly five spectral pixels
(as measured from the FWHM of calibration lamp lines), covering the range
5000--9200\AA.

The observations were done with two masks, each one with roughly 30
slits, with single integration times of 34 minutes each. Slits in the central
part of the mask were placed over the large gravitational arcs and other
candidate lensed objects. The remaining slits were positioned over bright
galaxies in the GMOS field.

Redshifts for galaxies with absorption lines were determined using the
cross-correlation technique \citep{T&D} as implemented by the software {\sc
RVSAO} \citep{rvsao}. We have used as templates the spectra of the NGC galaxies
1700, 1426, 3096, 4087, 4472, and 4751, observed with the 2.5-m 
Du~Pont telescope of the Las Campanas Observatory, and a synthetic spectrum 
built from  the stellar library of \citet{jacoby}.

All these templates were observed and reduced by Rodrigo Carrasco. More
details can be found in \citet{carrasco}. Only object-template matches with
the correlation coefficient parameter $R>3$ were considered successful. 
Typical formal errors in the radial velocities are smaller than 30~\kms.

For galaxies having prominent emission lines, the redshifts were also
determined directly from these lines, but these measurements were adopted only
when the velocity determinations by cross-correlation were doubtful ($R<5$). In
all cases, the differences between the two determinations were both results are
smaller than 150~\kms.  In total, we obtained 44 reliable redshifts. The
resulting redshifts for the cluster galaxies are summarized in Table
\ref{spectra}, and for the other galaxies in Table \ref{spectra_ncl}.

Objects are identified using their J2000.0 coordinates, in the same fashion as 
\citet{k&cohen}, so that C088\_3712 has coordinates R.A. =
07$^h$32$^m$08.$^s$8, DEC. = $+31\degr37\arcmin12\arcsec$.  The prefix ``C" 
in the name means that the galaxy is a cluster member, 
otherwise the prefix is a ``G".

\begin{deluxetable}{lcccrcl}
\tablewidth{0pt}
\tablecaption{Spectral data for Abell~586 galaxies. \label{spectra}}									   
\tablehead{																   
\colhead{(1)} & \colhead{(2)} & \colhead{(3)}& \colhead{(4)} & \colhead{~(5)} & 							   
\colhead{(6)} & \colhead{(7)}\\ 													   
\colhead{Name} & \colhead{RA (2000)} & \colhead{DEC (2000)} &\colhead{z} &								   
\colhead{~R} & \colhead{r$^\prime$ (AB Mag.)}												   
& \colhead{Emission lines}}														   
\startdata
C088\_3712                  & 07 32 08.81& 31 37 12.5& 0.1686 &   7.7 &  18.82 & H$\beta$, [OIII], [OI], [NII], H$\alpha$, [SII] \\		   
C093\_3835                  & 07 32 09.33& 31 38 35.1& 0.1687 &   6.7 &  20.12 & \nodata \\						   
C094\_3542                  & 07 32 09.39& 31 35 42.1& 0.1739 &   6.2 &  19.75 & \nodata \\						   
C107\_3616                  & 07 32 10.68& 31 37 24.8& 0.1744 &  13.7 &  18.84 & \nodata \\						   
C107\_3725                  & 07 32 10.74& 31 36 16.5& 0.1728 &   7.4 &  18.86 & H$\beta$, [OIII], [OI], [NII], H$\alpha$, [SII] \\		   
C118\_3731                  & 07 32 11.79& 31 37 31.2& 0.1699 &  12.9 &  18.34 & \nodata \\						   
C123\_3605                  & 07 32 12.35& 31 36 04.7& 0.1665 &   6.5 &  20.62 & \nodata \\						   
C127\_3828                  & 07 32 12.74& 31 38 27.8& 0.1655 &   6.0 &  20.86 & \nodata \\						   
C131\_3619~\tablenotemark{a}& 07 32 13.13& 31 36 18.6& 0.1597 &\nodata&  20.14 & H$\beta$, [OIII], [HeI], [OI], [NII], H$\alpha$, [SII] \\	   
C137\_3845                  & 07 32 13.70& 31 38 44.5& 0.1714 &  12.8 &  18.87 & \nodata \\						   
C144\_3717                  & 07 32 14.42& 31 37 17.3& 0.1730 &  13.0 &  18.04 & \nodata \\						   
C156\_3838                  & 07 32 15.56& 31 38 37.7& 0.1767 &  10.4 &  19.65 & \nodata \\						   
C157\_3723                  & 07 32 15.65& 31 37 22.7& 0.1802 &  14.4 &  18.53 & \nodata \\						   
C163\_3727                  & 07 32 16.32& 31 37 27.2& 0.1698 &  10.8 &  19.75 & \nodata \\						   
C169\_3839                  & 07 32 16.86& 31 38 39.3& 0.1683 &  14.4 &  18.54 & \nodata \\						   
C171\_3650~\tablenotemark{b}& 07 32 17.07& 31 36 50.0& 0.1738 &   5.9 &  18.23 & H$\beta$, [OIII], [OI], [NII], H$\alpha$, [SII] \\		   
C176\_3744                  & 07 32 17.56& 31 37 43.8& 0.1658 &   5.5 &  20.41 & \nodata \\						   
C177\_3857                  & 07 32 17.69& 31 38 57.3& 0.1679 &  13.0 &  18.06 & \nodata \\						   
C186\_3846~\tablenotemark{a}& 07 32 18.64& 31 38 46.5& 0.1673 &\nodata&  20.96~\tablenotemark{c} & H$\beta$, [OIII], [NII], H$\alpha$, [SII] \\
C187\_3846                  & 07 32 18.71& 31 38 46.0& 0.1595 &   4.9 &  20.96~\tablenotemark{c} & \nodata \\				   
C197\_3726                  & 07 32 19.74& 31 37 26.3& 0.1754 &   7.2 &  21.70 & \nodata \\						   
C223\_3818                  & 07 32 22.27& 31 38 17.9& 0.1720 &  16.3 &  19.02 & \nodata \\						   
C231\_3752                  & 07 32 23.08& 31 37 51.9& 0.1700 &  13.1 &  18.59 & \nodata \\						   
C233\_3800                  & 07 32 23.26& 31 38 00.3& 0.1780 &   6.2 &  20.68 & \nodata \\
C257\_3708                  & 07 32 25.75& 31 37 07.7& 0.1758 &   8.8 &  19.61 & \nodata \\						   
C266\_3652                  & 07 32 26.61& 31 36 52.2& 0.1666 &   4.4 &  21.38 & \nodata \\
C268\_3631                  & 07 32 26.77& 31 36 31.0& 0.1714 &  14.0 &  20.03 & \nodata \\						   
C273\_3618                  & 07 32 27.29& 31 36 17.9& 0.1715 &   8.0 &  19.58 & \nodata \\						   
C276\_3652                  & 07 32 27.57& 31 36 51.5& 0.1725 &  21.3 &  18.52 & \nodata \\						   
C281\_3813                  & 07 32 28.14& 31 38 12.6& 0.1686 &   8.8 &  17.67 & \nodata \\						   
C291\_3557~\tablenotemark{a}& 07 32 29.11& 31 35 56.5& 0.1710 &\nodata&  20.52 & [OIII], H$\alpha$, [SII]  \\				   
\enddata		      														   
\tablenotetext{a}{Redshift measured from emission lines.}										   
\tablenotetext{b}{This galaxy presents H$\alpha$ in emission with narrow and broad
($> 2000$ \kms) components, being probably a type 1 Seyfert galaxy. This object
is also the source of the point-like X-ray emission seen near the Southwest
corner of Figure \ref{AGN_xray}.}	    															   
\tablenotetext{c}{The images of these galaxies overlap, and the magnitude in
the table corresponds to the sum of both images.}													   
\end{deluxetable}															   

\begin{deluxetable}{lcccrcl}														   
\tablewidth{0pt}															   
\tablecaption{Spectral data for non-cluster members in the field of Abell~586.								   
\label{spectra_ncl}}															   
\tablehead{																   
\colhead{(1)} & \colhead{(2)} & \colhead{(3)}& \colhead{(4)} & \colhead{~(5)}
& \colhead{(6)} & \colhead{(7)}\\													   
\colhead{Name} & \colhead{RA (2000)} & \colhead{DEC (2000)} & \colhead{z} &								   
\colhead{~R} &\colhead{r$^\prime$ (AB Mag.)}
& \colhead{Emission lines}}
\startdata																   
G137\_3635	     	        & \phn07 32 13.67& 31 36 35.2& 0.2203 & 14.1  & 19.64 & \nodata \\
G147\_3845	     	        & \phn07 32 14.72& 31 38 45.1& 0.2123 &  7.5  & 18.78 & [NII], H$\alpha$ \\				   
G181\_3850	     	        & \phn07 32 18.07& 31 38 49.9& 0.3050 &  5.3  & 18.66 & \nodata \\					   
G194\_3816~\tablenotemark{a,c}  & \phn07 32 19.40& 31 38 15.8& 0.6093 &\nodata& 18.86 &  [OII], H$\beta$, [OIII] \\			   
G199\_3734~\tablenotemark{a,b,c}& \phn07 32 19.92& 31 37 33.6& 1.4302 &\nodata& 21.87 & [OII]	\\					   
G208\_3842~\tablenotemark{a,b}  & \phn07 32 20.76& 31 38 41.7& 0.8973 &\nodata& 21.89 & [OII]	\\					   
G243\_3809	     	        & \phn07 32 24.29& 31 38 08.7& 0.2453 &  3.6  & 22.25 & \nodata \\					   
G257\_3837                      & \phn07 32 25.69& 31 38 37.1& 0.3050 &  7.4  & 18.54 & \nodata \\					   
G283\_3604~\tablenotemark{a,b}  & \phn07 32 28.33& 31 36 04.2& 0.8563 &\nodata& 21.06 & [OII]	\\  
G288\_3703~\tablenotemark{a}    & \phn07 32 28.78& 31 37 03.3& 0.1921 &\nodata& 20.56 & [OIII], [NII], H$\alpha$ \\			   
G298\_3552	     	        & \phn07 32 29.79& 31 36 23.3& 0.1912 &  4.3  & 19.21 & [NII], H$\alpha$ \\	 			   
G298\_3623	     	        & \phn07 32 29.80& 31 35 51.6& 0.2141 &  8.1  & 19.31 & \nodata \\					   
G305\_3734~\tablenotemark{a}    & \phn07 32 30.52& 31 37 34.1& 0.0596 &\nodata& 19.16 & H$\beta$, [OIII], [NII], H$\alpha$, [SII] \\   	   
\enddata																   
\tablenotetext{a}{Redshift measured from emission lines.}										   
\tablenotetext{b}{Despite the presence of only one emission line, [OII], the
redshift of these galaxies has been confirmed by the presence of strong
absorption lines, namely CaII H and K, MgII and FeII.}
\tablenotetext{c}{Strong lensing features.}												   
\end{deluxetable}															   
\subsection{Chandra X-ray observation and data reduction \label{sec:chandra}}

Abell 586 was observed in September 2000 by the \textit{Chandra} satellite in a
single 11.83~ksec exposure with ACIS-I camera (observation ID 530, P.I. Leon
Van Speybroeck). The data were taken in Very Faint mode with a time resolution
of 3.24~sec. The CCD temperature was $-120^{\circ}$C. The data was reduced
using CIAO version 3.0.1\footnote{\texttt{http://asc.harvard.edu/ciao/}}
following the Standard Data Processing, producing new level 1 and 2 event
files.

The level 2 event file was further filtered, keeping only \textit{ASCA} grades
\footnote{The grade of an event is a code that identifies which pixels, within the three 
pixel-by-three pixel island centered on the local charge maximum, are above 
certain amplitude thresholds. The so-called ASCA grades, in the absence of 
pileup, appear to optimize the signal-to-background ratio.
\texttt{http://cxc.harvard.edu/}}
0, 2, 3, 4 and 6. We checked that no afterglow was present and applied the Good
Time Intervals (GTI) supplied by the pipeline. No background flares were
observed and the total livetime is 10.04~ksec.

We have used the CTI-corrected ACIS background event files (``blank-sky''),
produced by the ACIS calibration
team\footnote{\texttt{http://cxc.harvard.edu/cal/Acis/WWWacis\_cal.html}},
available from the calibration data base (CALDB). The background events were 
filtered, keeping the same grades as the source events, and then were 
reprojected to match the sky coordinates of Abell 586 ACIS observation.

We restricted our analysis to the range [0.3--8.0~keV], because above 8.0~keV,
the X-ray emission is largely background-dominated.

\subsubsection{X-ray Imaging}

\begin{figure}[!htb]
    \centering
    \includegraphics[width=\columnwidth]{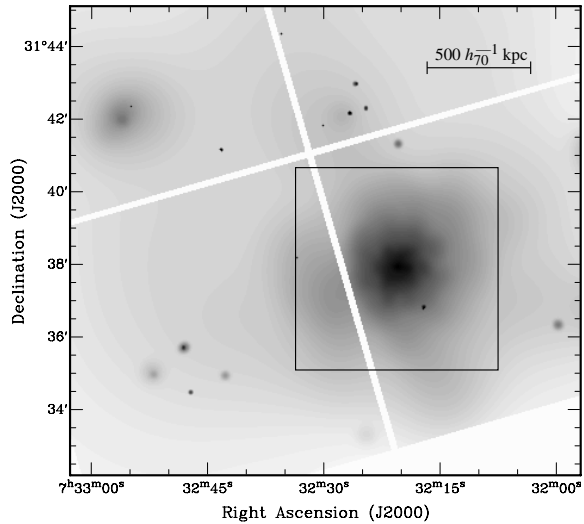}
    \caption{\textit{Chandra} ACIS-I X-ray adaptively smoothed image of Abell
    586 in the [0.3--8.0~keV] band, corrected by the exposure map and binned
    by a factor of 4. The center of the cluster is on the I3 chip (bottom right).
    There are also a number of point sources and a faint, extended source
    on the I0 chip (top left). The square ($5.5^{\prime}$ on a side) centered
    on the cluster shows the GMOS field of view.
    \label{fig:raw_smooth}}
\end{figure}

We have constructed an adaptively smoothed image in the [0.3--8.0~keV] band
using the \textsc{csmooth} tool from CIAO. The exposure map was generated by
the script \textsc{merge\_all}, where we have calculated the spectral weights,
needed for the instrument map, using the cluster total spectrum, i.e., the
spectrum obtained inside a circle concentric with the cluster, with a radius of
5~arcmin (about the GMOS field of view). We first smoothed the raw image, then
the exposure map, using the scale map produced by the smoothing of the raw
image, and finally divided the smoothed raw image by the smoothed exposure map.
The result is shown in Figure~\ref{fig:raw_smooth}.

The X-ray cluster emission can be detected up to 5~arcmin ($0.87 h_{70}^{-1}$
Mpc) and is fairly symmetric. There is a strong point source at 78~arcsec from
the center toward the Southwest that we identify with an active galaxy
C171\_3650 (cf. Table~\ref{spectra}).

The absence of significant substructure in the X-ray map suggests that the last
major merger with clusters with masses larger than 1/4
the Abell 586 mass
occurred on a timescale longer than the cluster relaxation time. 
Roughly, the relaxation time is of the order of the dynamical time,
$\tau_{d} \sim (G
\overline{\rho})^{-1/2}$ where $\overline{\rho} \sim 340 \Omega_{M} \rho_{c}$
inside the virial radius for a $\Lambda$CDM cosmology. Thus, $\tau_d  \sim 4
h_{70}^{-1}$Gyr and we suggest that the last major merger was at  least 4~Gyr
ago. Such a rough estimate usually agrees with $N$-body  simulation results
\citep[e.g.][]{Roettiger98,Rowley04}.

\subsubsection{X-ray Spectroscopy}

For the spectral analysis, we have computed the weighted redistribution and
ancillary files (RMF and ARF) using the tasks \textsc{mkrmf} and
\textsc{mkwarf} from CIAO. These tasks take into account the extended nature of
the X-ray emission. Background spectra were constructed from the blank-sky
event files and were extracted at the same regions (in detector coordinates) as
the source spectra that we want to fit.

The spectral fits were done using \textsc{xspec}~v11.3, restricting to the
range [0.3--8.0~keV]. The X-ray spectrum of each extraction region was modeled
as being produced by a single temperature plasma and we employed the
\textsc{mekal} \citep{Kaastra,Liedahl} model. The photoelectric absorption --
mainly due to neutral hydrogen -- was computed using the cross-sections given
by \cite{Balucinska}, available in \textsc{xspec}.

\begin{figure}[!htb]
    \centering
    \includegraphics[width=\columnwidth]{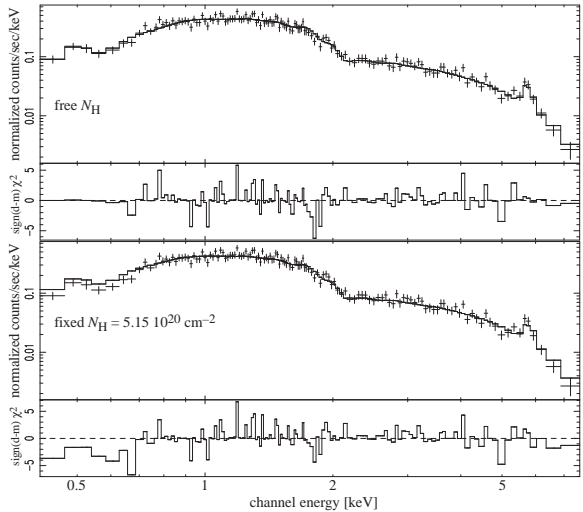}
    \caption{\textit{Chandra} ACIS-I X-ray spectrum extracted in the central
    1.4~arcmin ($245 h_{70}^{-1}\,$kpc) superposed with a \textsc{mekal}
    plasma spectrum. \textsl{Top:} fit with $N_{\rm H}$ free. \textsl{Bottom:}
    fit with $N_{\rm H}$ fixed at the galactic value. Below each spectrum
    the residuals are shown expressed as the $\chi^{2}$ contribution of each
    binned energy channel.
    \label{fig:a586Mekalcirc}}
\end{figure}

The overall spectrum was extracted within a circular region of 1.4~arcmin 
(243~kpc) centered on the X-ray emission peak. It was re-binned with the
\textsc{grppha} task, so that there are at least 30 counts per energy bin. This
radius was chosen because almost all emission is in this region, we can avoid
the CCDs gaps, and we have all the spectrum extracted in a single ACIS-I CCD.
Figure~\ref{fig:a586Mekalcirc} shows the overall spectrum fitted to an absorbed
\textsc{mekal} plasma.

A single-temperature gas fits well the overall spectrum; the reduced $\chi^{2}$
is $\chi^{2}/$d.o.f.$= 132.7/133$. The temperature thus obtained is $kT =
7.8_{-0.8}^{+1.0}$ keV (90\% confidence level), in agreement with previous
estimates;  the metal abundance (hereafter, metallicity) is $Z =
0.51_{-0.17}^{+0.18}~ Z_{\odot}$ and the hydrogen column density is $N_{\rm H}
= 9.1_{-1.3}^{+1.4} \times 10^{20}$ cm$^{-2}$. The metallicity is higher  than
the average value of $1/3 Z_{\odot}$ for clusters of galaxies.

The value of $N_{\rm H}$ obtained from the spectral fit is above the  galactic
value \citep{Dickey} which is $N_{\rm H} = 5.15 \times 10^{20}$  cm$^{-2}$ for
the region  of Abell 586. Not considering the uncertainties of the galactic
$N_{\rm H}$ measurements, this excess is significant at a $3\sigma$ level. Such
a level of significance may indicate that we are indeed detecting an X-ray
absorption excess. The origin of such an absorption is controversial \citep[and
references therein]{Allen00}, the most promising hypotheses being very cool
molecular gas and/or dust grains that survives in the intracluster medium
(ICM).

If we fix $N_{\rm H}$ at the galactic value, we obtain a much worst fit, 
$\chi^{2}/$d.o.f.$= 157.6/134$ and a higher temperature, $kT = 
9.8_{-1.0}^{+1.3}$keV [this anti-correlation between the temperature and 
hydrogen column density is well-known, cf. \citet{Pislar}].
As can be seen in the residual plot of figure \ref{fig:a586Mekalcirc},
this higher value of $\chi^{2}$ is mainly due to a poor fit at low energies
($E < 0.7$keV).  At high energies ($E > 6.5$ keV), the fit is also a bit
poorer than when $N_{\rm H}$ is considered as a free parameter.

\subsubsection{Flux and luminosity}

Table~\ref{tbl:Sumario} summarizes the non-absorbed fluxes and luminosities
obtained inside the central 1.4~arcmin field. We give the results in ``soft''
and ``hard'' bands, as well as the rest-frame bolometric luminosity (computed
by extrapolating the plasma emissivity from 10~eV to 100~keV).

\begin{table*}[htb]
\centering 
\caption{Non-absorbed fluxes and luminosities
in different energy bands. The energy band limits are given in keV.
\label{tbl:Sumario}}
\tabcolsep=0.75
\tabcolsep
\begin{tabular}{c c c  c c c}
\hline 
flux    &   flux     & flux   &  $L_{X}$   & $L_{X}$     & $L_{X}$\\
\relax [0.5--2.0] & [2.0--10.0] & bolom. & [0.5--2.0] & [2.0--10.0] & bolom.\\
\hline
2.8   &    5.4     &  11.3  &   2.08     &   4.38      &  9.18 \\
\hline
\end{tabular} 
\begin{flushleft}
\vspace{-1ex} Note: flux units are $10^{-12}$erg~cm$^{-2}\,$s$^{-1}$.\\ 
Luminosity units are $10^{44} \, h^{-2}_{70} \,$erg~s$^{-1}$.
\end{flushleft}
\end{table*}

Using the empirical relation $L_{X}$--$T_{X}$ obtained by \citet{Xue}, the
bolometric X-ray luminosity (converted to $H_{0}=50\,$km~s$^{-1}$~Mpc$^{-1}$)
implies $kT = 6.7 \pm 0.7$, slightly cooler but in agreement (within the error
bars) with the spectroscopically determined temperature.

\section{Data analysis}

\subsection{Dynamics of galaxies}

\begin{figure}[!htb]
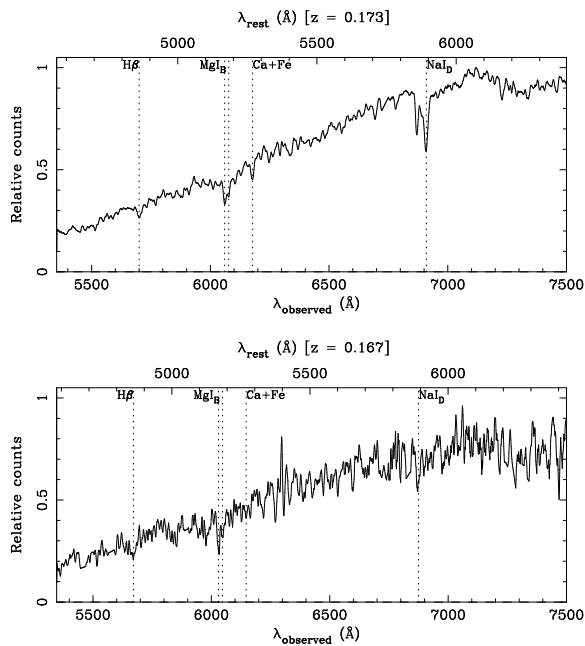

\centerline{\includegraphics[width=\columnwidth]{f3a.eps}}
\vskip 0.3cm
\centerline{\includegraphics[width=\columnwidth]{f3b.eps}}
\caption{{Examples of spectra of cluster member galaxies.
The upper and lower panels shows, respectively,  the spectra of the galaxy
with the best correlation coefficient (C276\_3652) and the worst one
(C266\_3652). Both spectra have not been flux calibrated and have been smoothed
using a boxcar filter with 6.8 \AA~(5 pixels) of length, for the sake of
clarity.} \label{example_spectrum}}
\end{figure}

\begin{figure}[!htb]
\centerline{\includegraphics[width=\columnwidth]{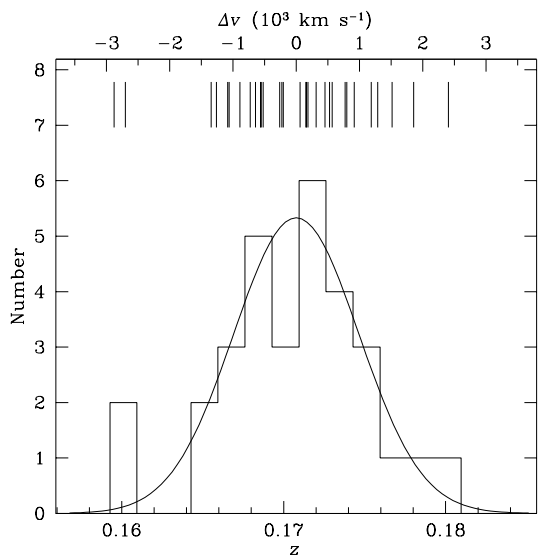}}
\caption{Redshift distribution for the 31 galaxies in Table \ref{spectra} with
$0.16 < z < 0.18$. The bin size is 500 \kms. The tick-marks on the top of the
panel represent individual galaxy velocities. The continuous line is a Gaussian
function with bi-weighted location and scale values equal to
$\langle z\rangle = 0.1708$ and $\sigma = 1161$ \kms,
respectively. \label{velhist}}
\end{figure}

We obtained accurate redshifts for forty-four galaxies in the GMOS field of
Abell 586, whose properties are summarized in Tables \ref{spectra} and
\ref{spectra_ncl}. Thirty-one of them, with redshifts between 0.16 and 0.18,
constitute our spectroscopic sample of cluster galaxies. In Figure
\ref{example_spectrum} we show examples of these spectra, and in Figure
\ref{velhist} their velocity distribution. The redshifts of the remaining
galaxies have either $z < 0.07$ or $z>0.19$.

We estimated the systemic redshift and the velocity dispersion of the cluster 
using the bi-weight estimators of the ROSTAT program \citep{beers}.  The
resulting mean redshift is $\langle z\rangle = 0.1708 \pm 0.0006$ and the
velocity dispersion is $\sigma = 1161\pm196$ \kms. 

This value of $\langle z \rangle$ compares very well with the value obtained by
\citet{bottini}  using only 7 radial velocities of galaxies with very
small projected distances to the BCG, $0.170 \pm 0.001$. His determined
velocity dispersion, $313\pm 70$ \kms, is much smaller than our value; this is
not unexpected given his small sample.

There is a noticeable gap in the velocity histogram of $\sim 1500$ \kms (in the
cluster rest frame) between the two galaxies at $z\sim 0.160$ and the others.
Excluding these two galaxies from our sample, the inferred values are
$\langle z\rangle = 0.1711\pm0.0006$
and $\sigma = 977\pm130$ \kms. However, we consider that these galaxies
should be kept in our sample because of the following
reasons.  Firstly, these galaxies are not excluded by a 3-sigma
clipping selection. Secondly, the velocity histogram becomes too asymmetric 
after exclusion of the two galaxies (the ROSTAT asymmetry index  increases
from  0.48 to 0.96).  Finally, because the difference between the BCG velocity
and the systemic velocity which is normally small in cD clusters
\citep{quintana&lawrie}, becomes larger, increasing from 121 to 221 \kms\, in
the cluster rest-frame.

We also verified the effect on the velocity dispersion caused by the 6
emission line galaxies, concluding that their removal from the calculation does
not cause any appreciable change in $\sigma$.   

Despite the internal robustness of the velocity dispersion found here, a word
of caution is warranted. All galaxies used in this analysis lie within a
radius of $570\, h_{70}^{-1}\,$kpc from the cluster center ($\sim 0.4 R_{\rm
vir}$; see Section \ref{sec:x-ray}). In order to understand more clearly the
dynamics of this cluster, a larger set of velocities obtained over a larger field
is required, as has been clearly demonstrated by, e.g., \citet{oliver}.

\subsection{Weak Gravitational Lensing}

\subsubsection{Galaxy shape measurements}

To estimate the cluster mass distribution through weak lensing, it is necessary
to measure accurately the ellipticity of the background galaxies, which includes
the effects of distortions introduced by the cluster gravitational potential.

The determination of galaxy shapes was performed using the method described in
detail by \citet{cypriano}. Here we only summarize the main steps of this
procedure.

The program \textsc{im2shape} \citep{im2shape} is the basic tool we used for shape
measurements. This program uses Bayesian methods to fit astronomical images as
a sum of two-dimensional Gaussian functions with elliptical bases. Each of them is
fully defined by six
parameters: two Cartesian center coordinates, ellipticity, position angle, 
semi-major axis size and the amplitude. Moreover, \textsc{im2shape} deconvolves the
fitted result by using a PSF, given also in terms of a sum of two-dimensional Gaussians, thus
recovering an unbiased and accurate estimation of the object's shape.

\textsc{im2shape} was first used to map the PSF over the whole GMOS field 
of  view, by determining the shape of unsaturated stars, which have been 
selected based on their FWHM. For the stars, we prevented
\textsc{im2shape} to do any deconvolution by using a Dirac's delta
function as the input PSF.

>From the original sample of stars we kept only those 
that actually sample the local PSF. This has been done through a
sigma-clipping process, where stars with too deviant ellipticities or position angles
were removed. The remaining stars map a PSF which is nearly
constant over the field, with average ellipticity of 0.056 (or 5.6\%)
and major axis oriented nearly East-West.

The next step is to run \textsc{im2shape} for galaxies. As galaxy images have more complex
shapes than stars, they were modeled by a sum of two two-dimensional elliptical Gaussians
with the same center, ellipticity and position angle. 
For each galaxy, the input PSF was calculated using their five closest stars.

\subsubsection{Sample selection}

Once we have measured the galaxy shapes, we need to identify
the background galaxies, which can be used as probes of the cluster shear
field. These galaxies constitute what we call the weak-lensing sample. 
Since their redshift is unknown, they were selected by their magnitudes and colors
(when available). We have included in the weak-lensing sample 
only galaxies fainter than r$^\prime = 23.5$ and with (r$^\prime$
- i$^\prime$) colors not closer than 0.2 mag to the cluster red sequence.
These criteria are adequate because the number density radial profile 
of the weak-lensing sample is nearly flat and does not decrease with
increasing radius, as one would expect if there was significant contamination 
by cluster members.

In order to select a good quality weak-lensing sample, we kept only 
objects having errors on the tangentially projected ellipticities (with
respect to the cluster center) smaller than 0.35. The resulting weak-lensing 
sample has 276 galaxies; the faintest has r$^\prime = 25.6$ mag., and the
average magnitude in this sample is r$^\prime = 24.8$.

\subsubsection{Mass distribution}

\begin{figure}[!htb]
\centerline{\includegraphics[width=\columnwidth]{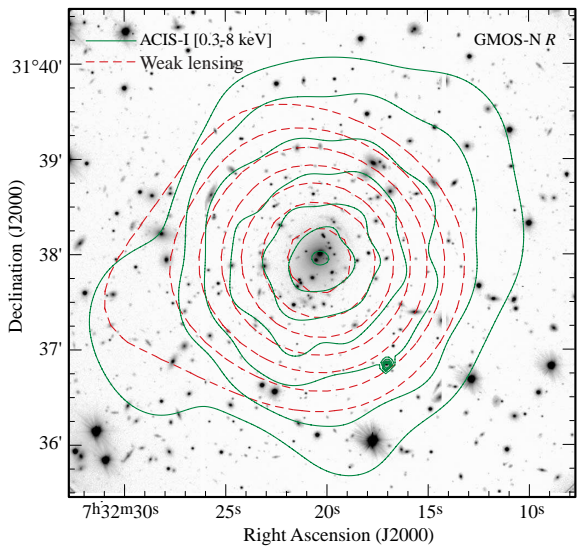}}
\caption{GMOS image of the central region of Abell 586 superimposed with
logarithmically spaced X-ray isophotes (continuous lines) and weak lensing
reconstructed mass density (dashed lines). The X-ray point source near the
southwest corner is the Seyfert 1 galaxy C171\_3650. \label{AGN_xray}}
\end{figure}

The information on the shape of these galaxies (and the corresponding errors)
was used to feed the software \textsc{LensEnt} \citep{sarah1,phil} which, based
on a maximum entropy method,
reconstructed the projected mass density distribution of Abell~586.
The resulting mass contours can be viewed in Figure
\ref{AGN_xray}. This map has been smoothed with a two-dimensional Gaussian 
with FWHM = 2.0\arcmin, that maximizes the likelihood of the 
reconstructed mass density given the data.

The mass distribution presented in
this figure, showing a single mass clump associated with the BCG, is
qualitatively very similar to the X-ray emission. This strongly suggests that the
central region of Abell 586 is indeed in dynamical equilibrium. It is also
worth noticing an extension towards the SE direction of the cluster, which is
present both in the X-ray and weak-lensing contours.
Our map is also in qualitative agreement with the weak-lensing mass map
produced by \citet{dahle}, despite of a small offset in the peak position
which, however, is within the resolution of both maps.

\subsubsection{Radial shear profile}

Very often the shear profile is fitted by a singular
isothermal profile \citep[SIS; e.g.][]{mellier99}.
This model is very convenient because it has a single
parameter, the Einstein radius, $\theta_E$, and allows direct comparisons with
results already obtained for this and other clusters. Besides, it offers a 
rough approximation when 
only the central part of the cluster can be accessed, as in the present 
case. In this model, $\theta_E$ is directly related to the 
cluster velocity
dispersion by the expression:
\eq
\theta_E = 4 \pi {\sigma^2 \over c^2} \beta \, , 
\eeq 
where $\beta$ is the ratio of $D_{ls}$, the angular diameter distance
between the lens and the source, over $D_s$, the angular diameter 
distance to the source.

We adopt a simple parametric model for the mass distribution, the singular
isothermal profile (SIS). More complex models are not appropriate,
because our data has a rather high noise level 
due to the small number of galaxies to probe the shear field.
Additionally, this kind of profile is widely used in studies of
this type, allowing a simple comparison with other clusters.

To estimate the average $\beta$ of our sample we have used the catalog of
photometric redshifts for the HDF of \citet{soto}. From this catalog, we
selected a sample with the same bright limit and average magnitude as the
present weak lensing sample. This catalog, however, does not have r$^\prime$
magnitudes; thus we have used I magnitudes, adopting the color  (r$^\prime - $
I) = 1.55. This color is typical for a Sbc galaxy at $z=0.8$ \citep{Fuku95}. This
procedure resulted in an average value  $\langle\beta \rangle = 0.678$,
corresponding to an average redshift of $0.78\pm 0.05$.

\begin{figure}[!htb]
\centerline{\includegraphics[width=\columnwidth]{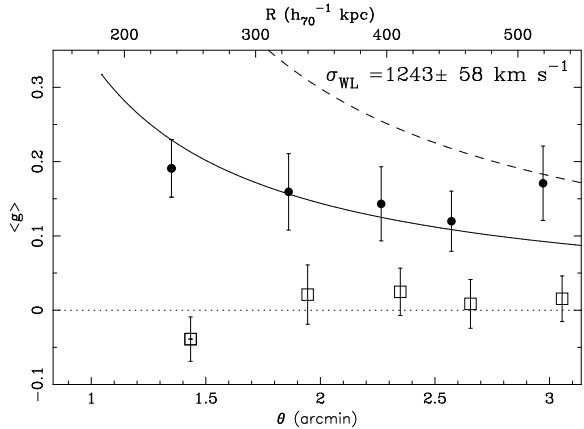}}
\caption{The shear profile for the cluster Abell 586. The filled symbols
correspond to average ellipticities of the faint background galaxies projected
tangentially to the cluster center. Each point represents  the average of at
least 55 galaxies of the weak-lensing sample. The open squares are the same,
but for ellipticities projected in a direction 45\degr\ relative to the center.
The solid line shows the best-fit SIS profile. The dashed line shows the
\citet{dahle} best SIS model ($\sigma$ = 1680 \kms).
\label{shear_prof}}
\end{figure}

The best-fit shear profile (Figure \ref{shear_prof}) gives an Einstein radius
of $30\pm 3$\arcsec, which translates into  $\sigma = 1243\pm 58$ \kms. The
uncertainty of 58 \kms is the statistical error of the fitting. It is important
to mention that the determination of $\langle\beta \rangle$ is  the major
source of systematic uncertainties on the derived velocity dispersion and thus
on absolute mass determinations through weak-lensing. For instance, if we adopt
the color of a typical Sab or Scd galaxy, instead of a Sbc, the resulting
velocity dispersion changes about $\pm 50$ \kms.

The value of the velocity dispersion we obtain here is significantly
smaller than the one found by \citet{dahle}, $1680^{+160}_{-170}$ \kms. It is
difficult to figure out the reasons of this discrepancy. One possible reason
is that  Dahle et al. have used the approximation $g \sim \gamma$ instead of $g = \gamma
/ (1-\kappa)$ used on the present work. Here, $g$ is the distortion, which is 
directly related to background-galaxy ellipticities, $\gamma$ is the
gravitational shear,  and $\kappa$ is the convergence or the projected mass
density normalized with respect to a critical mass density:
\begin{equation}
\kappa \equiv {\Sigma \over \Sigma_c} =
\Sigma \left({c^2 \over 4 \pi G} {1 \over D_l~\beta}\right)^{-1} \, .
\end{equation}

For strong lensing clusters like Abell~586, the assumption implicit in the
approximation used by Dahle et al., i.e. $\kappa\sim 0$, is too strong, mainly
for the cluster central regions. However, adopting this approximation in our
sample, we get $\sigma = 1353$ \kms.  This valu is closer to the Dahle et al.
result, but it is still significantly smaller. It is worth mentioning that the
observations employed by them have similar depth to our observations. Dahle et
al. observations were done with the 2.56-m NOT telescope with exposures of
5.4 ks in both $V$ and $I$ filters. After scaling and summing both exposures,
the resulting imaging is about the same of the present 1.2 ks r$^\prime$
imaging with the 8.1-m Gemini-N telescope. Both detectors have also a similar
field-of-view. In terms of image quality, however, our data is  better. Dahle
et al. reported a seeing FWHM of 1.0 and 0.8 arcsec for their $V$ and $I$
images, respectively, whereas for our data this value is 0.7 arcsec.

\subsubsection{Total mass} 

The mass profile can be directly measured from the shear data using the technique
called \textit{aperture mass densitometry} \citep[AMD;][]{fah94}.
This method is based on the single assumption that the surface mass density
is circularly symmetric.

The AMD actually measures the difference between the mass density inside a
given radius
and the ring between this radius and a maximum radius of reference ($r_{max}$):
\begin{equation}
\label{amdeq}
M_{AMD} (<r) = \pi~r^2\Sigma_c \, \zeta(r,r_{max}) \, ,
\end{equation}
where $\zeta$ is defined by:
\begin{eqnarray}
\zeta(r_1,r_{max}) & = \bar\kappa(r<r_1) -  \bar\kappa(r_1<r<r_{max})\nonumber \\
 & = {2 \over 1 - (r_1/r_{max})^2}
\int^{r_{max}}_{r_1} {g_t(r) ~d \ln{r}} \, ,
\end{eqnarray}
where $g_t$ is the tangentially projected distortion, $r_1$ the radius in which
the mass is measured, and $\bar\kappa$ the mean convergence.

We choose $r_{max} = 150\,$arcsec (437~kpc at $z=0.17$), which is the largest
radius fully contained within the field of view. Therefore, the total mass 
inside 422~kpc ($145\arcsec$)
\footnote{largest $r_1$ so that there is at least 20 galaxies of the
weak-lensing sample inside the annulus $r_1<r<r_{max}$.} estimated by AMD is
$(4.3 \pm 0.7)\times 10^{14}\, M_\odot$. The mass profile can be seen, together
with X-ray mass profiles, in Section \ref{totalmass}.

>From equation (\ref{amdeq}) it can be seen that the mass measured using the AMD
method depends on $\Sigma_c$ and thus on $\beta$. Therefore the same systematical
uncertainties related to the poorly known redshift distribution of the
background sources, as previously discussed, applies here.

\subsubsection{Strong-Lensing Features}

\begin{figure}[!htb]
\centerline{\includegraphics[width=\columnwidth]{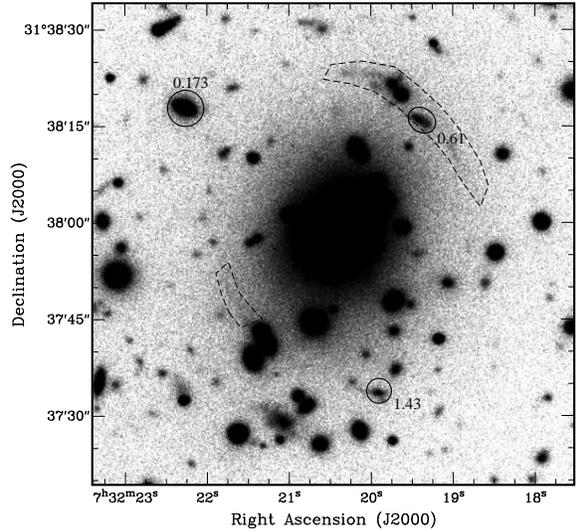}}
\caption{Optical image of central region of Abell 586. Objects with redshifts
measured by us inside this field are marked. The dashed lines surround the
gravitational arcs. This image shows a square region of 1.2\arcmin\ on a
side.\label{central}}
\end{figure}

The central region of Abell 586 (see Figure~\ref{central})
shows several low surface brightness structures
oriented tangentially to the cluster center, most of them with
(g$^\prime$--i$^\prime$) colors up to 0.5 mag bluer than cluster ellipticals with
similar brightness.

A prominent giant arc, already reported by Dahle et al. (2002), can be seen in 
the Northwest direction, distant 22 arcsec of the BCG center.
There are several high surface brightness galaxies superimposed on this arc.
At the opposite side of the central galaxy another giant arc, although fainter,
can be appreciated in Figure~\ref{central} at 20 arcsec from the
cluster center. Unfortunately, we could not determine the redshift of these
arcs and, therefore, could not confirm whether they are multiple images of the same
source. 

We succeeded, however, to measure the redshifts of two arclets in the vicinity
of the BCG, at 26.8 arcsec and 19.7 arcsec from the BCG center, whose
orientations are nearly tangential to the direction of the cluster center. The
spectra of these objects present emission and interstellar absorption
lines, both typical of late-type galaxies. Their measured redshifts are 1.43
and 0.61, respectively (cf. Figure \ref{central}). No other objects with colors
similar to those of these arclets were found, so no additional candidates 
for multiple images could be identified.

In the absence of multiple gravitational images it is not possible to model the
cluster potential using the position of the arcs  \citep[like in][]{kneib96}.
However, a rough estimate of the cluster mass inside the region enclosed by the
arclets can be obtained assuming  that they put a limit on the
position of the Einstein radius. Under this assumption, we obtain, from the
higher and lower redshift arclets, $\sigma$ equal to 1056 \kms\ and 998
\kms, respectively. 

\begin{figure}[!htb]
\centerline{\includegraphics[width=\columnwidth]{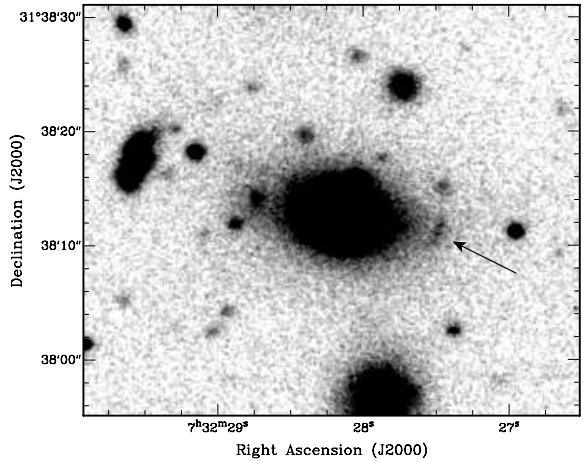}}
\caption{Image of the galaxy C281\_3813, showing an arc-like object.
\label{gal_arc}}
\end{figure}

This cluster shows another strong-lensing-like feature that deserves be
mentioned.  
In Figure \ref{gal_arc} we present a close-up of the spiral galaxy and cluster
member C281\_3813, that is at 1.7\arcmin\ from the cluster center. What is
remarkable in this figure is the presence of an arc-like object (indicated by an
arrow). If this object has indeed a gravitational origin, it would be an uncommon
example of strong-lensing by a late-type galaxy.

\subsection{X-ray \label{sec:x-ray}}

\subsubsection{X-ray brightness profile}

\begin{figure}[!htb]
   \centering
   \includegraphics[width=\columnwidth]{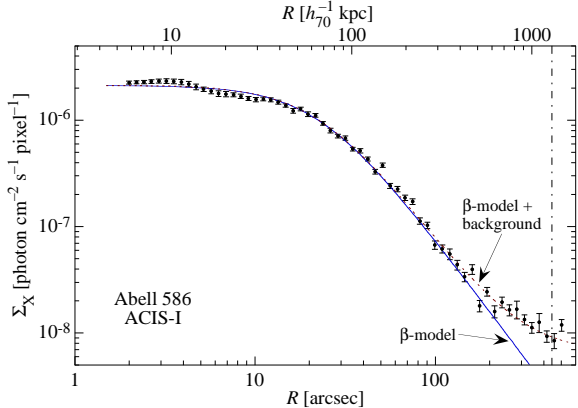}
    \caption{X-ray brightness profile. The full line shows the best fit
    $\beta$-model with $\beta = 0.518 \pm 0.006$ and $R_{c} = 23.1 \pm
    0.6$~arcsec ($67 \pm 2\, h_{70}^{-1}$kpc); the dotted line shows the
    $\beta$-model plus a constant background brightness (which was actually
    fitted). The vertical dotted-dashed line indicates the virial radius (cf.
    Section~\ref{sec:virialR}).
   \label{fig:A586_BrightProf}}
\end{figure}

The X-ray brightness profile of Abell 586 was obtained with the STSDAS/IRAF
task \textsc{ellipse}. We have used the exposure map corrected image in the
[0.3--8.0 keV] band, binned by a factor 4 (one X-ray image pixel has 2~arcsec). We
have masked the CCDs gaps and source points. The brightness profile, shown in
Figure~\ref{fig:A586_BrightProf}, could be measured up to 500~arcsec ($1.46\,
h_{70}^{-1}$Mpc) from the cluster center.

Using the $\beta$-model \citep{Cavaliere76} to describe the surface brightness
radial profile,
\begin{equation}
     \Sigma_{X}(R) = \Sigma_{0} \left[ 1 +
     \left(\frac{R}{R_{c}}\right)^{2}\right]^{-3\beta/2} \, ,
\end{equation}
a least-squares fitting gives $\beta = 0.518 \pm 0.006$ and $R_{c} = 23.1 \pm
0.6$~arcsec ($67 \pm 2\, h_{70}^{-1}$kpc). If we assume that the gas is
approximately isothermal and distributed with spherical symmetry, there is a
simple relation between the brightness profile and the gas number density,
$n(r)$, i.e.,
\begin{equation}
     n(r) = n_{0} \left[ 1 +
     \left(\frac{r}{r_{c}}\right)^{2}\right]^{-3\beta + 1/2} \, ,
     \label{eq:gasDens}
\end{equation}
where $R_{c} = r_{c}$ (capital indicates projected 2D coordinates, lower case
indicates 3D coordinates).

In order to estimate the central electronic density, $n_{0}$, we integrate the
bremsstrahlung emissivity along the line-of-sight, in the central region, and
compare with the flux obtained by spectral fitting of the same region. We
obtain thus $n_{0} = (18.4 \pm 1.5) \times 10^{-3}\,$cm$^{-3}$.

\subsubsection{Temperature profile}

\begin{figure}[htb]
    \centering
    \includegraphics[width=\columnwidth]{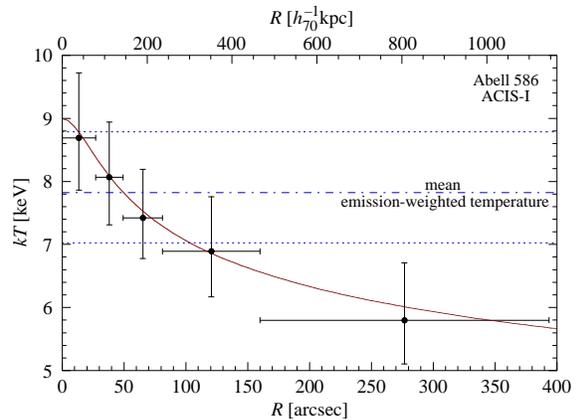}
    \caption[]{Temperature profile. The error bars are $1\sigma$ confidence
    level. The horizontal dotted-dashed line is the mean temperature inside
    1.4~arcmin, $kT = 7.8^{+1.0}_{-0.8}$ keV, the horizontal dotted lines
    correspond to 90\% confidence level of the mean temperature. The full line
    shows the best-fit of the polytropic model (see text).
    \label{fig:Perfil_kT_poly}}
\end{figure}

We have computed the radial temperature profile using concentric circular
annuli. For each annulus, defined by approximately the same number of counts
(2000 counts, background corrected), a spectrum was extracted and fitted
following the method described above, except that the hydrogen column density
and metallicity were kept fixed at the mean best-fit value found inside
1.4~arcmin (i.e., $N_{\rm H} = 9.1_{-1.3}^{+1.4} \times 10^{20}$cm$^{-2}$ and
the metallicity $Z = 0.5~Z_\odot$). Figure~\ref{fig:Perfil_kT_poly} shows the
temperature profile.

Since the temperature profile shows a clear gradient, we have tried a simple 
parametric form of the ICM temperature using a polytropic equation of state. 
It is not clear whether a polytropic model describes well the ICM. Indeed, 
\citet{Irwin99} and \citet{deGrandi02} argue that this is not a good 
description of the  gas temperature profile in clusters. However, hydrodynamic
simulations  \citep{Suto98}, theoretical models
\citep[e.g.,][]{Komatsu01,dosSantos02}  and some observations
\citep[e.g.,][]{Markevitch99,Finoguenov01a}  suggest that the gas may  be
described by a polytropic model, with polytropic index $\gamma$ roughly in  the
range $1.1 \la \gamma \la 1.2$.

Therefore, we have fitted a polytropic temperature profile:
\begin{equation}
    T(r) = T_{0} \left[ 1 + \left(\frac{r}{r_{c}}\right)^{2} 
    \right]^{-3\beta (\gamma-1)/2} \, ,
    \label{eq:TempProfile}
\end{equation}
where $r_{c}$ and $\beta$ are the values obtained with the $\beta$-model fitting of the
brightness surface profile, and $T_{0}$ is the central  temperature.

A standard least-square fit of Eq.~(\ref{eq:TempProfile}) with only two free 
parameters, $T_{0}$ and $\gamma$, yields a rather good fit: $T_{0} = 8.99 \pm 
0.34$ keV and $\gamma = 1.10 \pm 0.03$, cf. Figure~\ref{fig:Perfil_kT_poly}.
This  value agrees with those found by \citet{Finoguenov01b} and, being well
below  5/3 (the ideal gas value), suggests that the gas may be in adiabatic
equilibrium  \citep[see, eg.,][Sect.~5.2]{Sarazin88}.

We note that this cluster does not present any sign of cooling in the very
central part, at $R \approx 70\, h_{70}^{-1}\,$kpc, the smallest radius where
we can extract a meaningful spectrum and measure the temperature. Either we
lack the resolution to detect an eventual drop in temperature or the
intracluster gas is not cooling. Since the central cooling-time is roughly
\begin{equation}
    t_{\rm cool} \approx 5.8 \times 10^{9} \frac{T_{\rm keV}^{1/2}}{n_{3}} 
    \approx 10^{9} \, \mbox{years} \, ,
\end{equation}
then, if the gas is indeed not cooling, something must be heating it
\citep[as it was already realized for some clusters, e.g.,][]{Peterson01}.
Heating by cluster merging seems improbable, given the apparent spherical
symmetry of the X-ray
emission. Other possibilities, like heating by AGN, thermal
conduction, etc., may be playing a role in the energy budget of this cluster
\citep[e.g.,][]{Markevitch03}.

However, we may be simply not detecting an eventual drop in temperature because
we lack the resolution. Using a sample of 20 clusters, \citet{Kaastra04} show
that the radius ($r_{c}$ in their paper) where the temperature drops in
cooling-flow clusters is, with 2 exceptions, smaller than $70
h_{70}^{-1}\,$kpc.

\subsubsection{Gas and total masses \label{totalmass}}

\begin{figure}[htb]
    \centering
    \includegraphics[width=\columnwidth]{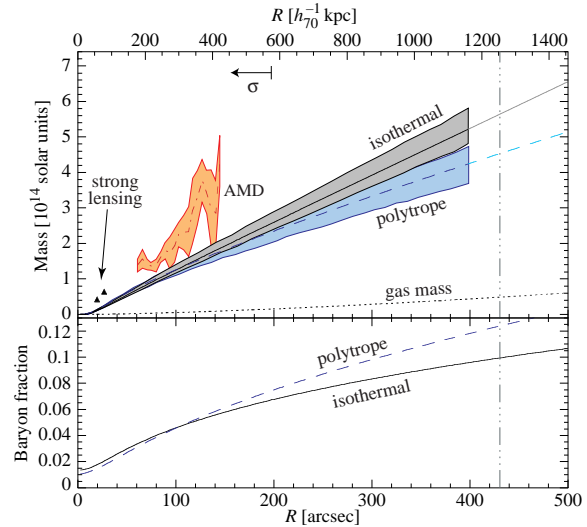}
    \caption{\textsl{Top:} X-ray dynamical mass profile assuming isothermal
    profile (full line), polytropic profile (dashed), 
    weak-lensing mass profile using the AMD method
    (small dashed) and strong-lensing mass estimates (triangles). The 
    shaded areas represents 1-$\sigma$ uncertainties. The arrow on the top
    of the panel marks the radius that contains all the galaxies used for the
    dynamical analysis.    
    \textsl{Bottom:}
    Baryon fraction for isothermal (full line) and polytropic (dashed)
    temperature profiles. The vertical line indicates the virial radius.
    \label{fig:Mass_BarionA586}}
\end{figure}

We estimate the gas mass simply by integrating the density given by 
Eq.~(\ref{eq:gasDens}), assuming spherical symmetry which, in this case, seems 
a good approximation (cf. Figure~\ref{AGN_xray}).
The integrated gas mass is shown in the top panel of
Figure~\ref{fig:Mass_BarionA586}.

Having the temperature profile, we also compute the ``X-ray'' dynamical mass
(i.e., the dynamical mass determined from an X-ray observation, not to be
confused with the mass of the X-ray emitting gas). Supposing hydrostatic
equilibrium and polytropic temperature profile, the dynamical mass is given by:
\begin{equation}
    M(r) = \frac{3 kT_{0} \beta \gamma r_{c}}{G \mu m_{\rm p}} \left(
    \frac{r}{r_{c}} \right)^{3} \left( 1 + 
    \left[\frac{r}{r_{c}}\right]^{2}\right)^{-1 - \frac{3}{2}(\gamma-1) 
\beta} \, .
    \label{eq:masspoly}
\end{equation}
If $\gamma = 1$, we have the usual
isothermal mass profile. Here, $\mu\, m_{p}$ is the mean mass per particle
($\mu = 0.6$ for a fully ionized gas with primordial He abundance) and $m_{\rm
p}$ is the proton mass. The top panel of Figure~\ref{fig:Mass_BarionA586} shows
the dynamical mass profile estimated both with an isothermal ($\gamma = 1$) and
polytropic temperature profile  ($\gamma = 1.1$).

The gas mass at $R = 400''$ ($1.16\, h_{70}^{-1}$Mpc) is $0.43 \times 10^{14} 
M_{\odot}$. Depending on the assumed temperature profile, $M_{\rm dyn} = 5.3 
\times 10^{14} M_{\odot}$ (isothermal case) and $M_{\rm dyn} = 4.25 \times 
10^{14} M_{\odot}$ (polytropic case) within the same radius.

\subsubsubsection{Virial radius and baryon fraction\label{sec:virialR}}

We can now compute the radius corresponding to the density ratio $\delta =
\bar{\rho}(r_{\delta})/\rho_{\rm c}(z)$. For $\delta = 200$, we have the usual
$r_{200}$; for the SCDM and $\Lambda$CMD cosmological models, $\delta = 180$
and 340 correspond, respectively, to the virial radius
\citep[cf.][]{Lacey93,Bryan98}.

Taking into account the polytropic temperature profile, we use the formula in
the appendix of \citet{LimaNeto03} to compute the various $r_{\delta}$.
Table~\ref{tab:rvir} shows the results, where each column corresponds to a
density ratio and the lines with ``poly'' and ``isoth'' refer to the
polytropic and isothermal temperature profiles, respectively. Then, assuming
a polytropic profile and the $\Lambda$CDM scenario, the virial radius is
$r_{\rm vir} = 1.26 h_{70}^{-1}\,$Mpc, which corresponds to $R = 430''$.

\begin{deluxetable}{ccccccc}
\tablewidth{0pt}
\tablecaption{Radius at different density contrast.
\label{tab:rvir}}
\tablehead{
\colhead{$\qquad\delta$:} & \colhead{180} & \colhead{200} & \colhead{340} & 
\colhead{500} & \colhead{2500} & \colhead{Model}}
\startdata
$r_{\delta}$: & 1.69 & 1.61& 1.26& 1.05 & 0.49 & poly \\
    arcsec       & 1.97 & 1.83& 1.40& 1.16 & 0.51 & isoth \\
\enddata \\[3pt]
Obs.: $r_{\delta}$ is in units of  $h_{70}^{-1}$Mpc.
\end{deluxetable}

Using the dynamical and gas mass profiles, we can derive the baryon fraction
profile, $f(r) \equiv M_{\rm baryon}/M_{\rm total}$, assuming that the bulk of
the baryons are in the ICM (i.e., $M_{\rm baryon} \approx M_{\rm
gas}$). The galactic contribution to the baryon mass is taken into account 
following \citet{White93} and \citet{Allen}:
\begin{equation}
    M_{\rm baryon} = M_{\rm gas} (1 + 0.16\, h_{70}^{0.5}) \, ,
    \label{eq:MassGalBaryon}
\end{equation}
where $M_{\rm gas}$ is the intracluster gas mass. 

At $R = 400$ arcsec, the baryon fraction is still rising. Its value at this point
depends on the assumed temperature profile: for the isothermal and polytropic
cases we have $f = 0.10$ and 0.12, respectively. The baryon fraction determined
here agrees with the usual values found in rich clusters
\citep[e.g.][]{WhiteFabian,Allen}.

At the virial radius, defined above, we have $M_{\rm gas} = 0.48 \times 10^{14}
M_{\odot}$, $M_{\rm dyn} = 4.53 \times 10^{14} M_{\odot}$, and $f = 0.12$
(assuming a polytropic temperature profile).

\section{Discussion}

The mass determinations resulting from the application of four
distinct techniques, based on different types of data, to the cluster Abell 586
are compared in this section  discussed within the context
of eventual deviations from a relaxed state.

In Table~\ref{tab:sigma}  are listed the velocity dispersions
either measured or deduced in this paper through strong-lensing, X-ray,
redshift survey, and weak-lensing methods, together with results of
other authors.

Except for the velocity dispersion determined by \citet{bottini} and
\citet{dahle}, all  estimations agree well within at least a 68\% confidence
level. This result is strengthened
when we compare the mass profiles provided by each technique
(see Figure~\ref{fig:Mass_BarionA586}),
although systematic
differences within 2$\sigma$ between the (model dependent) 
measurements are noticed.

\begin{deluxetable}{cll}
\tablewidth{0pt}
\tablecaption{Cluster velocity dispersion derived with different methods.
\label{tab:sigma}}
\tablehead{
\colhead{(1)} & \colhead{(2)} & \colhead{(3)} \\
\colhead{$\sigma$ [km~s$^{-1}$]} & \colhead{Method} & \colhead{Notes}}
\startdata
998--1056 & strong-lensing & $\theta_{E}$ 19.7\arcsec--26.8\arcsec, sources
$z=0.61$ and 
1.43\\[6pt]
$1015 \pm 500$ & X-ray luminosity\tablenotemark{a} & $L_{X} = 18 \times 10^{44}\,$erg~s$^{-1}$\\
$1174 \pm 130$ & X-ray temperature\tablenotemark{a} & $kT = 7.8$ keV \\
\textit{1050} $\pm$ \textit{350} & \textit{X-ray}\tablenotemark{a} & \textit{Allen (2000)} 
\\[6pt]
$1161 \pm 196$ & velocity distribution & 31 galaxies \\
\textit{313 $\pm$ 70} & \textit{velocity distribution} & \textit{7 galaxies~\citet{bottini}}\\[6pt]
$1243 \pm 58$ & weak lensing & \\
\textit{1680 $\pm$ 170} & \textit{weak lensing} & \textit{Dahle et al. (2002)} \\
\enddata \\[3pt]
\tablenotetext{a}{For X-ray data we have used the empirical relations $L_{X}$--$\sigma$
and $T_{X}$--$\sigma$ from \citet{Xue}.}
\tablecomments{Italic entries are quoted results from the literature.}
\end{deluxetable}

Recently, \citet{cypriano} compared mass estimates obtained through different
techniques  for a sample of 24 X-ray luminous clusters with $z<0.3$ and with
homogeneous weak-lensing determinations. Adopting the criteria that the
agreement, or disagreement, between dynamical (velocity dispersion and X-ray)
and non-dynamical (weak-lensing) mass estimators in a particular cluster is
indicative that the cluster is relaxed, or not, we have
found that clusters with ICM temperatures above $\sim$8.0~keV show strong
evidences of dynamical activity, while cooler clusters tend to be nearly
relaxed.

Our study of this cluster suggests that Abell 586 is a well relaxed object that
has not experienced a major merger in the  last few Gyr; notice that its ICM
temperature is just below the upper $\sim$8.0~keV limit found for quasi-relaxed
systems, so that this indicator may have a valid predictive character.

Gravitational arcs are often found in clusters that are not relaxed. This
behavior can be understood not only because the more massive clusters are in
general young structures, but also because the presence of substructures and
other features associated with dynamical activity, enhance the strong-lensing
cross-section, as shown, for example, in ray-tracing simulations by 
\citet{bartelmann}.

Indeed, a recent study by \citet{graham} of a sample of 10 strong-lensing
clusters, selected by X-ray luminosities, using HST and high quality X-ray data,
concluded that only 30\% of them can be classified as truly relaxed clusters.
Actually, the Cypriano et al. criteria successfully predicts the
dynamical state of 80\% of this sample.

Given that cluster inner mass profiles are often determined
by strong-lensing analysis \citep[e.g.,][]{tyson,A383,sand1,kneib03,sand2}, and
possibly the majority of these systems are non-relaxed, then, the disagreement 
between these observed profiles and the theoretical dark matter profiles,
derived from relaxed halos found
in numerical simulations \citep[e.g.][]{NFW,ghigna,El-Zant}, might be due to the
fact that  the physical state of the observed and modelled systems
are not consistent.

The identification and detailed mass reconstruction of a representative sample
of clusters like Abell 586, using several techniques like gravitational
lensing, X-ray emission and galaxy velocities, is a promising way towards a
better understanding of the behavior of baryonic and dark matter components in
the  center of galaxy clusters. 

\section{Conclusion}

Using optical data taken with the 8-m Gemini-North telescope and available 
\textit{Chandra} X-ray data for the A586 galaxy cluster, we have derived its
mass distribution and content using a  multi-technique analysis. Our main
results can be summarized as follows:

\begin{enumerate}

\item Radial velocity measurements for 31 cluster galaxies resulted on
a systemic redshift of $\langle z\rangle= 0.1708\pm0.0001$ and a
velocity dispersion $\sigma = 1161\pm196$ \kms.

\item We detected weak gravitational shear, whose best fit 
through an isothermal mass profile gives $\sigma = 1243\pm58$ \kms.

\item We identified a system of gravitational arcs and
determined the redshifts for two arclets
($z=$ 0.61 and 1.43) belonging to this system. 

\item We determined the mass distribution in the central
region of the cluster through two techniques: weak-lensing and X-ray emission;
they are found to be very similar and having almost circular geometry.

\item The ICM gas is distributed very smoothly; it has a mean temperature of
$7.8^{+1.0}_{-0.8}\,$keV and a mean metallicity of $0.51^{+0.18}_{-0.17}
Z_\odot$, both values being slightly higher than the averages  reported in
the literature for rich clusters.

\item The gas temperature profile is well described by a polytropic model with
$\gamma = 1.1$.

\item The cluster virial radius is approximately $1.3 h_{70}^{-1}$ Mpc, and the
gas and dynamical mass within this radius are  $M_{\rm gas} = 0.48 \times
10^{14} M_{\odot}$ and  $M_{\rm dyn} = 4.53 \times 10^{14} M_{\odot}$; the
baryon fraction at the same radius is $f = 0.12$, assuming a polytropic
temperature profile.

\end{enumerate}

The ensemble of our observational results, derived with different techniques
and wavelenght ranges, indicate consistently that A586 is a massive cluster,
characterized by a velocity dispersion that is in the range 1000--1250 \kms.
Several pieces of evidence suggest that this cluster is dynamically well
relaxed, namely: $i$) The near circular mass and X-ray luminosity distribution,
both concentric with the BCG; $ii$) The agreement, uncertainties taken into
account, between dynamical (X-ray, galaxies velocity dispersion) and
non-dynamical (gravitational lensing) mass estimators; $iii$) An ICM
temperature profile well described by a polytrope with index $\gamma = 1.1$.

As a final remark, it is interesting to note that Abell 586 is found to follow
the \cite{cypriano} criteria to diagnose the dynamical state of luminous X-ray
clusters. Its ICM temperature is just below the upper $\sim$8.0~keV limit 
claimed by Cypriano et al. for quasi-relaxed systems.

\acknowledgments ESC (CNPq-Brazil fellow), GBLM and LS acknowledge financial
support from CNPq and FAPESP.  JPK acknowledgs support from CNRS and Caltech.
LEC thanks support from Fondecyt grant No. 1040499. We also thank Hugo Capelato
for help on the dynamical analysis. 
We are grateful to the staff of the Gemini Observatary for undertaking the
queue observing for this project in an efficient manner.

This work is partially based on observations obtained at the Gemini
Observatory, which is operated by the Association of Universities for Research
in Astronomy, Inc., under a cooperative agreement with the NSF on behalf of the
Gemini partnership: the National Science Foundation (United States), the
Particle Physics and Astronomy Research Council (United Kingdom), the National
Research Council (Canada), CONICYT (Chile), the Australian Research Council
(Australia), CNPq (Brazil) and CONICET (Argentina). Proposal ID's:
GN-2001B-Q-15 and GN-2002B-Q-5 


\begin{thebibliography}{}

\bibitem[Allen(1998)]{allen} Allen, S. W. 1998
\mnras, 296, 392

\bibitem[Allen(2000)]{Allen00} Allen, S. W. 2000 \mnras, 315, 269

\bibitem[Allen et al.(2002)]{Allen} Allen S.W., Schmidt R.W. \& Fabian A.C., 
2002, MNRAS 334, L11

\bibitem[Balucinska-Church \& McCammon(1992)]{Balucinska} Balucinska-Church M.
\& McCammon D., 1992, ApJ 400, 699

\bibitem[Bartelmann, Steinmetz \& Weiss(1995)]{bartelmann} Bartelmann, M.,
Steinmetz, M. \& Weiss, A. 1995, \aap, 291, 1 

\bibitem[Beers, Flynn \& Gebhartd(1990)]{beers} Beers, T. C., Flynn  K. \&
Gebhardt 1990, \aj, 100, 32

\bibitem[Bertin \& Arnouts(1996)]{SEx} Bertin, E. \& Arnouts, S. 1996, A\&AS,
117, 393

\bibitem[Bottini(2001)]{bottini} Bottini, D. 2001, \aj, 121, 1294

\bibitem[Bridle et al.(1998)]{sarah1}Bridle, S. L., Hobson, M. P., Lasenby, A.
N. \& Saunders, R. 1998, \mnras, 299, 895

\bibitem[Bridle et al.(2002)]{im2shape} Bridle, S., Kneib, J.-P., Bardeau, S.
\& Gull, S.F., 2002, in:``The shapes of Galaxies and their Dark Halos'', Yale
Cosmology Workshop, 28-30 May 2001, World Scientific

\bibitem[Bryan \& Norman(1998)]{Bryan98} Bryan G.L. \& Norman M.L., 1998,
\apj,  495, 80

\bibitem[Carrasco, Mendes de Oliveira \& Infante(2005)]{carrasco} Carrasco, E.
R. D., Mendes de Oliveira, C. L. \& Infante, L. 2005, \textit{in preparation}

\bibitem[Cavaliere \& Fusco-Femiano(1976)]{Cavaliere76} Cavaliere A. \&
Fusco-Femiano R., 1976, A\&A, 49, 137

\bibitem[Cohen \& Kneib(2002)]{k&cohen} Cohen, J. G. \& Kneib, J.-P. 2002,
\apj, 573, 524

\bibitem[Cypriano et al.(2004)]{cypriano} Cypriano, E. S., Sodr\'e Jr., L.,
Kneib, J.-P. \& Campusano, L. E. 2004, ApJ, 613, 95 

\bibitem[Czoske et al.(2002)]{oliver} Czoske, O., Moore, B., Kneib, J.-P. \&
Soucail, G. 2002, \aap, 386, 31 

\bibitem[Dahle et al.(2002)]{dahle} Dahle, H., Kaiser, N., Irgens, R. J.,
Lilje, P. B. \& Maddox, S. J. 2002, \apjs, 139, 313

\bibitem[De Grandi \& Molendi(2002)]{deGrandi02} De Grandi, S. \& Molendi, S., 
2002, \apj, 567, 163

\bibitem[Dickey \& Lockman(1990)]{Dickey} Dickey, J.M. \& Lockman, F.J., 1990,
Ann. Rev. Ast. Astr., 28, 215

\bibitem[Dos Santos \& Dor\'e(2002)]{dosSantos02} Dos Santos S. \& Dor\'e O., 
2002, A\&A, 383, 450

\bibitem[Ebeling et al.(1998)]{BCS} Ebeling, H., Edge, A. C., Bohringer, H.,
Allen, S. W., Crawford, C. S., Fabian, A. C., Voges, W. \& Huchra, J. P. 1998,
\mnras, 301, 881

\bibitem[El-Zant et al. (2004)]{El-Zant} El-Zant, A. A., Hoffman, Y., Primack,
J., Combes, F. \& Shlosman, I. 2004, \apjl, 607, 75

\bibitem[Fahlman et al.(1994)]{fah94} Fahlman, G., Kaiser, N., Squires, G. \&
Woods, D. 1994, \apj, 437, 56

\bibitem[Fern\'andez-Soto, Lanzetta \& Yahil(1999)]{soto} 
Fern\'andez-Soto, A., Lanzetta, K. M. \& Yahil, Amos 1999, \apj, 513, 34

\bibitem[Ferrari et al.(2003)]{ferrari03} Ferrari C., Maurogordato, S., Cappi,
A. \& Benoist, C., 2003, \aap, 399, 813

\bibitem[Finoguenov et al.(2001a)]{Finoguenov01a} Finoguenov A., Reiprich T.H.
\& B\"ohringer H., 2001a, A\&A, 368, 749

\bibitem[Finoguenov et al.(2001b)]{Finoguenov01b} Finoguenov A., Arnaud M. \&
David L.P., 2001b, \apj, 555, 191

\bibitem[Fort \& Mellier(1994)]{FM94} Fort, B. \& Mellier Y. 1994, \aapr, 5,
239

\bibitem[Fukugita, Shimasaku \& Ichikawa(1995)]{Fuku95} Fukugita, M.,
Shimasaku, K. \& Ichikawa, T. 1995, PASP, 107, 945

\bibitem[Fukugita et al.(1996)]{sloan} Fukugita, M., Ichikawa, T., Gunn, J. E.,
Doi, M., Shimasaku, K. \& Schneider, D. P. 1996, \aj, 111, 1748

\bibitem[Girardi \& Mezzeti(2001)]{girardi} Girardi, M  \& Mezzeti, M. 2001, 
\apj, 548, 79

\bibitem[Ghigna et al.(1998)]{ghigna} Ghigna, S., Moore, B., Governato, F., 
George. L., Quinn, T. \& Stadel J. 1998, \mnras, 300, 146

\bibitem[Hattori, Kneib \& Makino(1999)]{hkm99} Hattori, M., Kneib, J.-P. \&
Makino, N. 1999, Progress of Theoretical Physics, 133, 1 

\bibitem[Hoekstra(2003)]{hoekstra} Hoekstra, H. 2003, \mnras, 339, 1155

\bibitem[Hook et al.(2002)]{GMOS} Hook, Isobel, Allington-Smith, J. R., Beard,
S., Crampton, D., Davies, R., Dickson, C. J., Ebbers, A., Fletcher, M.,
Jorgensen, I., Jean, I., Juneau, S., Murowinski, R., Nolan, R., Laidlaw, K.,
Leckie, B., Marshall, G.E., Purkins, T., Richardson, I., Roberts, S., Simons,
D., Smith, M., Stilburn, J., Szeto, K., Tierney, C. J., Wolff, R. \&  Wooff, R.
2002, SPIE, 4841, Power Telescopes and Instrumentation into the New Millennium

\bibitem[Irwin et al.(1999)]{Irwin99} Irwin J.A., Bregman J.N. \& Evrard A.E.,
1999, \apj, 519, 518

\bibitem[Jacoby, Hunter \& Christian(1984)]{jacoby} Jacoby, G. H., Hunter, D.
A. \& Christian, C. A. 1984, \apjs, 56, 257 

\bibitem[Kaastra \& Mewe(1993)]{Kaastra} Kaastra, J.S., Mewe, R., 1993,  A\&AS
97, 443  

\bibitem[Kaastra et al.(2004)]{Kaastra04} Kaastra, J. S., Tamura T., Peterson
J. R., Bleeker, J. A. M., Ferrigno, C., Kahn, S. M., Paerels, F. B. S.,
Piffaretti, R. \& Branduardi-Raymont G., B\"ohringer H., 2004, A\&A 413, 415

\bibitem[Kauffmann et al.(1999)]{kauffmann99} Kauffmann G.,  Colberg, J. M.,
Diaferio, A. \& White, S. D. M., 1999, MNRAS, 303, 188

\bibitem[Komatsu \& Seljak(2001)]{Komatsu01} Komatsu, E. \& Seljak, U., 2001,
\mnras, 327, 1353

\bibitem[Kurtz \& Mink(1998)]{rvsao} Kurtz, M. J. \& Mink, D. J. 1998, \pasp,
110, 934

\bibitem[Kneib et al.(1996)]{kneib96} Kneib, J.-P., Ellis, R. S., Smail, I.,
Couch, W. J. \& Sharples, R. M. 1996 \apj, 471, 643

\bibitem[Kneib et al.(2003)]{kneib03} Kneib, J.-P., Hudelot, P. Ellis, R. S.,
Treu, T., Smith, G. P., Marshall, P., Czoske, O., Smail, I. \& Natarajan, P.
2003, \apj, 598, 804

\bibitem[Lacey \& Cole(1993)]{Lacey93} Lacey C. \& Cole S., 1994, \mnras, 271,
676

\bibitem[Liedahl et al.(1995)]{Liedahl} Liedahl, D.A., Osterheld, A.L. \& 
Goldstein, W.H., 1995, \apjl, 438, 115

\bibitem[Lima Neto et al.(2003)]{LimaNeto03} Lima Neto, G. B., Capelato, H. V.,
Sodr\'e, L., Jr.\& Proust, D., 2003, A\&A 398, 31

\bibitem[Markevitch et al.(1999)]{Markevitch99} Markevitch, M., Vikhlinin, A., 
Forman, W. R. \& Sarazin, C. L., 1999, \apj 527, 545

\bibitem[Markevitch et al.(2003)]{Markevitch03} Markevitch, M., Vikhlinin, A.
\&  Forman, W. R., 2003, in: ``Matter and Energy in Clusters of Galaxies'', ASP
Conf.  Series 301, 37 (Eds. S. Bowyer, C.-Y. Hwang)

\bibitem[Marshall et al.(2002)]{phil} Marshall, P. J., Hobson, M. P., Gull, S.
F. \& Bridle, S. L. 2002, \mnras, 335, 1037 

\bibitem[Mellier(1999)]{mellier99} Mellier, Y. 1999, \araa, 37, 127

\bibitem[Metzger \& Ma(2000)]{metzgera697} Metzger, M. R. \& Ma, C.-P. 2000,
\aj, 120, 2879 

\bibitem[Metzeler et al.(1999)]{metzeler} Metzler, C. A., White, M., Norman \&
M., Loken, C. 1999, \apjl, 520, 9 

\bibitem[Miralda-Escud\'e \& Babul(1995)]{miralda}  Miralda-Escud\'e \& J.
Babul, A. 1995, \apj, 449, 18

\bibitem[Navarro, Frenk \& White(1997)]{NFW} Navarro, J. F., Frenk, C. S. \&
White, S. D. M. 1997, \apj, 490, 493

\bibitem[Tyson, Kochanski \& dell'Antonio(1998)]{tyson} Tyson, J. A.,
Kochanski, G. P. \& dell'Antonio, I. P. 1998, \apjl 498, 107

\bibitem[Peterson et al.(2001)]{Peterson01} Peterson, J.R., Kahn, S. M. \&
Paerels, F. B., S., et al., 2001,  \apj, 590, 207

\bibitem[Pislar et al.(1997)]{Pislar} Pislar, V., Durret, F., Gerbal, D., Lima
Neto, G. B. \& Slezak, E., 1997, A\&A, 322, 53

\bibitem[Proust et al.(2003)]{proust03} Proust, D., Capelato, H. V., Hickel,
G., Sodr\'e, L., Jr., Lima Neto, G. B. \& Cuevas, H., 2003, \aap, 407, 31

\bibitem[Quintana \& Lawrie(1982)]{quintana&lawrie} Quintana, H. \& Lawrie, D.
G. 1982, \aj, 87, 1

\bibitem[Roettiger et al.(1998)]{Roettiger98} Roettiger, K., Stone, J.M., 
Mushotzky, R.F., 1998, \apj, 493, 62

\bibitem[Rowley et al.(2004)]{Rowley04} Rowley, D.R., Thomas, P.A.,
Kay, S.T., 2004, \mnras, 352, 508

\bibitem[Sand, Treu \& Ellis(2002)]{sand1} Sand, D. J., Treu, T. \& Ellis, R.
S., 2002, \apjl, 574, 129

\bibitem[Sand et al.(2004)]{sand2} Sand, D. J., Treu, T. Smith, G. P. \& Ellis,
R. S., 2004, \apj,604, 88

\bibitem[Sarazin(1988)]{Sarazin88} Sarazin C.L., 1988, ``X-ray emissions from 
cluster of galaxies'', Cambridge Astrophysics Series

\bibitem[Smail et al.(1997)]{morphs} Smail, I., Ellis, R. S., Dressler, A.,
Couch, W. J., Oemler, A. Jr., Sharples, R. M. \& Butcher, H. 1997, \apj, 479,
70

\bibitem[Smith et al.(2001)]{A383} Smith, G. P., Kneib, J.-P., Ebeling, H.,
Czoske \& O. Smail, I.  2001 \apj, 552, 493

\bibitem[Smith et al.(2003)]{graham03} Smith, G. P., Edge, A. C., Eke, V. R.,
Nichol, R. C., Smail, I. \& Kneib, J.-P. 2003, \apjl, 590, 79 

\bibitem[Smith et al.(2004)]{graham} Smith, G. P., Kneib, J.-P., Smail, I.,
Mazzotta, P., Ebeling, H. \& Czoske, O., submitted to MNRAS, astro-ph(0403588)

\bibitem[Suto et al.(1998)]{Suto98} Suto, Y., Sasaki, S. \& Makino, N., 1998,
\apj, 509, 544

\bibitem[Tonry \& Davis(1979)]{T&D} Tonry, J. \& Davis, M. 1979, \aj, 84, 1511

\bibitem[Valtchanov et al.(2002)]{valtchanov} Valtchanov, I., Murphy, T.,
Pierre, M., Hunstead, R. \& L\'emonon, L. 2002, \aap, 392, 795

\bibitem[White et al.(1993)]{White93} White, S. D. M., Navarro, J. F. \&
Evrard, A. E., Frenk, C. S.,  1993, Nature 366, 429

\bibitem[White \& Fabian(1995)]{WhiteFabian} White, D. A. \& Fabian, A. C.,
1995, \mnras, 273, 72

\bibitem[White \& Rees(1978)]{white78} White, S. D. M. \& Rees, M. J., 1978, 
\mnras, 183, 341

\bibitem[White(2000)]{white2000} White, D. A. 2000, \mnras, 312, 663

\bibitem[Xue \& Wu(2000)]{Xue} Xue, Y.J. \& Wu, X.P., 2000, \apj, 538, 65

\end{thebibliography}
 \end{document}